\documentclass[aps,twocolumn,pra,superscriptaddress,nofootinbib,amsmath,amssymb,floats,floatfix,english]{revtex4-1}

\usepackage[T1]{fontenc}
\usepackage[latin9]{inputenc}
\usepackage{color}
\usepackage{graphicx}
\usepackage{dsfont}

\def\be{\begin{equation}}
\def\ee{\end{equation}}
\def\bea{\begin{eqnarray}}
\def\eea{\end{eqnarray}}

\begin{document}

\title{Two-body mobility edge in the Anderson-Hubbard model in three dimensions: \\
 Molecular versus scattering states}

\author{Filippo Stellin} 
\email{filippo.stellin@univ-paris-diderot.fr}
\affiliation{Universit\' e de Paris, Laboratoire Mat\' eriaux et Ph\' enom\`enes Quantiques, CNRS, F-75013, Paris, France}
\author{Giuliano Orso}
\email{giuliano.orso@univ-paris-diderot.fr}
\affiliation{Universit\' e de Paris, Laboratoire Mat\' eriaux et Ph\' enom\`enes Quantiques, CNRS, F-75013, Paris, France}



\begin{abstract}
Most of our quantitative understanding of disorder-induced metal-insulator transitions comes from numerical studies of simple noninteracting tight-binding models, like the Anderson model in three dimensions. An important outstanding problem is the fate of the Anderson transition in the presence of additional Hubbard interactions of strength $U$ between particles. 
Based on large-scale numerics, we compute the position of the  mobility edge for a system of two identical bosons or two fermions with opposite spin components.  
The resulting phase diagram in the interaction-energy-disorder space possesses a remarkably rich and counterintuitive structure, with multiple metallic and insulating phases. We show that this phenomenon originates from the molecular or scattering-like nature of the pair states available at given energy $E$ and disorder strength $W$. 
The  disorder-averaged density of states of the effective model for the pair is also investigated. 
Finally, we discuss the implications of our results for ongoing research on many-body localization.
\end{abstract}

\maketitle

\section{Introduction}
A central concept in the physics of disordered systems is  
Anderson localization~\cite{Anderson:LocAnderson:PR58},
namely the absence of wave diffusion in certain random media as a result of interference effects between the 
 multiple scattering paths generated by the  impurities. To date, this phenomenon has been reported for different 
  kinds of waves, including light waves in diffusive media~\cite{Wiersma:LightLoc:N97,Maret:AndersonTransLight:PRL06} or in 
  disordered photonic crystals~\cite{Segev:LocAnderson2DLight:N07,Lahini:AndersonLocNonlinPhotonicLattices:PRL08},
ultrasound~\cite{vanTiggelen:AndersonSound:NP08}, microwaves~\cite{Chabanov:StatisticalSignaturesPhotonLoc:N00} and 
atomic matter waves~\cite{Billy:AndersonBEC1D:N08,Roati:AubryAndreBEC1D:N08}, to cite a few. 
  
Being an interference effect, Anderson localization crucially depends on the spatial dimension of the system and the underlying symmetries  of the associated model, which determines its universality class. 
In the absence of magnetic fields and spin-orbit couplings,  
 the Hamiltonian of a quantum particle exhibits both  time-reversal  and spin-rotational symmetries and therefore belongs to the  orthogonal class~\cite{Altland:PRB1997}.    
 For an uncorrelated disorder, all wave-functions  are then exponentially localized in one and two dimensions. 
In three dimensions, however, the energy spectrum  contains one or more critical points, called mobility edges, separating localized from extended states. 
 At these points the system undergoes a metal-insulator phase transition, known as Anderson transition~\cite{Evers:AndersonTransitions:RMP08},
  which is characterized by universal critical exponents.
 Mobility edges have been reported~\cite{Kondov:ThreeDimensionalAnderson:S11,Jendrzejewski:AndersonLoc3D:NP12,Semeghini:2014}  in experiments with 
noninteracting ultracold atoms in three-dimensional (3D) speckle potentials. Analogous transition for light waves, despite several claims, have not yet been 
 unambiguously observed, mainly due to the vector character of light~\cite{Skipetrov:PRL2014}.  

 Anderson transitions are difficult to describe analytically and our  quantitative understanding relies heavily on numerics. The most studied example of a disordered system is a tight-binding model with random onsite energies,  known as the Anderson model. 
 In first quantization notation, the latter writes
\begin{equation}\label{Anderson3D}
\hat H^\textrm{sp}= -J \sum_{\langle \mathbf n, \mathbf m\rangle} |\mathbf m  \rangle  \langle \mathbf n| + \sum_{\mathbf n}V_\mathbf n |\mathbf n\rangle \langle \mathbf n|, 
\end{equation}
where $J$ is the tunneling rate between two nearest neighboring sites 
$\mathbf n$ and $\mathbf m$,
 while $V_{\mathbf n}$ are random variables denoting the local value of the disorder potential. For simplicity, the disorder is assumed to be spatially uncorrelated, $\langle V_\mathbf n V_{\mathbf n^\prime} \rangle= \langle V_\mathbf n^2\rangle \delta_{\mathbf n \mathbf n^\prime}$ and obeying a uniform on site distribution
\be\label{randombox}
P(V)=\frac{1}{W}\Theta(W/2-|V|),
\ee
where $\Theta(x)$ is the Heaviside function and $W$ is the disorder strength. 
The position of the mobility edge for the model (\ref{Anderson3D})
was first computed in Ref.~\cite{Bulka:ZPB1987} using transfer matrix techniques. These results, which extended previous work~\cite{McKinnonKramer:TransferMatrix:ZPB83} 
performed for zero energy of the particle, were instrumental to develop approximate semianalytical theories of the Anderson transition, 
including  the  self-consistent theory of localization~\cite{Vollhardt:SelfConsistentTheoryAnderson:92,Economou:PRB1984,Kroha:SelfConsistentTheoryAnderson:PRB90}. 

The Anderson model is currently investigated in three~\cite{Slevin:CriticalExponent:NJP14} and higher dimensions~\cite{Ueoka:2014,Tarquini:PRB2017} to pinpoint the precise position of the mobility edge and to provide accurate estimates of the universal critical exponents. 
The same model emerges from the discretization of the Schr\"odinger equation of a continuum system. In particular, approximating the  Laplacian  by a second order finite difference yields  Eq.~(\ref{Anderson3D}) with  $J=\hbar^2/(2m \Delta^2)$, where $m$ is the particle mass and $\Delta$ is the lattice spacing. 
This procedure has recently been applied to obtain precise estimates~\cite{Delande:MobEdgeSpeckle:PRL2014,Pilati:LevelStats:2015,Pasek:3DAndersonSpeckle:PRA2015,Pilati:3DAndersonSpeckle:2015,Pasek:PRL2017,Orso:SpinOrbit:PRL2017} for the position of the mobility edge of cold atoms in laser speckle potentials, taking into account both the spatial correlations and the specific onsite distribution of the disorder.  
 
A main topic of current research is many-body localization~\cite{Nandkishore2015,ALET2018498,Abanin:RMP2019}, namely 
the generalization of Anderson localization to disordered systems of \emph{interacting} quantum particles.  
Of particular interest are many-body mobility edges, namely critical points at finite energy density, separating the many-body localized phase at weak interaction from the metallic, ergodic, phase at strong interaction. Experimental evidence of such critical points has been reported~\cite{Schreiber:Science2015,Kondov:PRL2015,Choi1547,Rispoli:Nature2019} in experiments with ultra-cold atoms in disordered lattices, implementing either the fermionic or the bosonic Anderson-Hubbard model in various dimensions.  
From the theoretical side,  numerical studies  of systems 
with a finite density of particles have mainly focused on one-dimensional models~\cite{PhysRevB.75.155111,Andraschko:PRL2014,PhysRevB.91.081103,Mondaini:PRA2015,Reichl:PRA2016,Prelovsek:PRB2016,Zakrzewski:PRB2018,krause2019nucleation,yao2020manybody}, due to the high computational effort.   
The existence of many-body mobility edges in systems with space dimension larger than one is currently debated~\cite{Agarwal:Annalen2017}. Analytical 
arguments were given~\cite{DeRoeck:PRB2016,DeRoeck:PRB2017} suggesting that the many-body localized phase in the thermodynamic limit is inherently unstable against the formation of thermal bubbles. 
This prediction contrasts with  numerical investigations of
 two-dimensional disordered Hubbard models~\cite{WahlNatPhys2019,geiler2019manybody}, providing evidence for a many-body localized phase at strong disorder. Similar conclusions have also been reached for 2D models of spinless fermions with
 nearest-neighbor interactions~\cite{De_Tomasi_2019,Thomson:PRB2018} and of quantum dimers~\cite{thveniaut2019manybody}.

A second and complementary approach to interaction-induced  Anderson transitions 
focuses on  few-body systems, starting from the solution of the two-particle problem in the presence of disorder.  
The corresponding Hamiltonian can be written in second quantization as $\hat H=\hat H_0 + \hat U$, where
$\hat H_0=\hat H^\textrm{sp} \otimes  \hat{\mathds{1}} +\hat{\mathds{1}}  \otimes \hat H^\textrm{sp}$ is the noninteracting part and
\be\label{intro1}
\hat U=U\sum_{\mathbf m}|{\mathbf m},{\mathbf m}\rangle \langle {\mathbf m},{\mathbf m}|
\ee
 is the onsite Hubbard interaction  of strength $U$. 
For 1D systems, the problem of two-particle localization was first addressed by Shepelyanski~\cite{Shepelyansky:AndLocTIP1D:PRL94}. Using results from random matrix theory, he showed that, in the presence of disorder, two particles coupled via short-range interactions can spread over a distance 
much larger than the single-particle localization length, before being ultimately localized. This surprising effect has been confirmed by 
several numerical studies~\cite{Weinmann:PRL1995,vonOppen:AndLocTIPDeloc:PRL96,Frahm1999,Roemer:PhysicaE2001,Krimer:JETP2011,Dias:PhysicaA2014,Lee:PRA2014,Krimer:InterConnDisord2PStates:PRB15,Frahm:EigStructAL1DTIP16,Thongjaomayum:PRB2019,thongjaomayum2020multifractality} during the last 25 years, although the analytical formula describing the enhancement of the pair localization length at weak disorder is still debated. 
The localization properties of a one dimensional system of few (two, three) interacting bosonic atoms subject to a laser speckle disorder have  recently been addressed~\cite{Mujal:PRA2019}.

Anderson localization of few interacting photons states in a disordered chain has been discussed theoretically for both linear~\cite{Lee:PRA2014} and nonlinear~\cite{See:PRA2019} photonic lattices. Remarkably, signatures of interaction-induced delocalization have been recently observed~\cite{Roushan1175} experimentally in a chain of superconducting qubits simulating the disordered Bose-Hubbard model.  Quantum correlations in the dynamics of two interacting particles moving in a disordered lattice have also been 
investigated~\cite{Lahini:PRL2010,Chattaraj:PRA2016,Dariusz:PRA2017} with application to nonclassical light and ultracold atoms.

In Refs~\cite{Borgonovi:NonLinearity1995,Imry:CohPropTIP:EPL95} it was argued that all two-particle states  remain localized in one and two dimensions (although the pair localization length can be extremely large), whereas in three dimensions  an Anderson transition to a diffusive phase 
could occur even when all single-particle states are localized. 
These claims are in clear contrast  with subsequent numerical works~\cite{Ortugno:AndLocTIPDeloc:EPL99,Roemer1999}, providing evidence of 2D metal-insulator transitions 
 of the pair induced by the Hubbard interactions (although finite-size effects can be an important issue).

Based on large scale numerical  simulations, we recently investigated~\cite{Stellin:PRB2019} the two-particle problem in three dimensions, 
focusing on a pair with zero total energy, $E=0$.
We addressed the localization properties of the system by mapping the original Hamiltonian  onto an effective single-particle model [see Eq.~(\ref{integral}) below] describing the center-of-mass motion of the pair, following the lines of Ref.~\cite{Dufour:PRL2012}. We found that Anderson transitions of the pair were consistent with the orthogonal universality class, although the inclusion of irrelevant variables in the finite-size scaling analysis was crucial to obtain accurate results for the mobility edge.
Interestingly,  single-particle excitations in a disordered electronic system with Coulomb interaction have also been shown~\cite{Burmistrov:PRB2014} to undergo an Anderson transition which belongs to the noninteracting universality class.

In Ref.~\cite{Stellin:PRB2019} we derived the phase diagram  in the interaction-disorder plane for a pair with zero total energy, $E=0$. For a given value of the interaction strength $U$, we found a single critical disorder amplitude $W_c$ separating the extended states $(W<W_c)$ from the localized ones $(W>W_c)$.
Moreover, we showed that the metal-insulator transition for the pair occurs in a regime where all single-particle states are localized,
confirming that interactions favor the \emph{delocalization} of the pair, irrespective of their  attractive or repulsive nature. 
The opposite effect, that is interaction-induced \emph{localization} of the pair, is also possible. Indeed two particles can form attractively  or repulsively  bound states.
 For sufficiently strong interactions, so that $E\simeq U$, these states behave as point-like particles with reduced tunneling rate $2J^2/|U|$. As a consequence, they tend to localize already in the presence of a very weak disorder, as previously observed~\cite{Dufour:PRL2012} for
1D quasiperiodic lattices.
 \begin{figure}
 	\includegraphics[width=0.360\columnwidth]{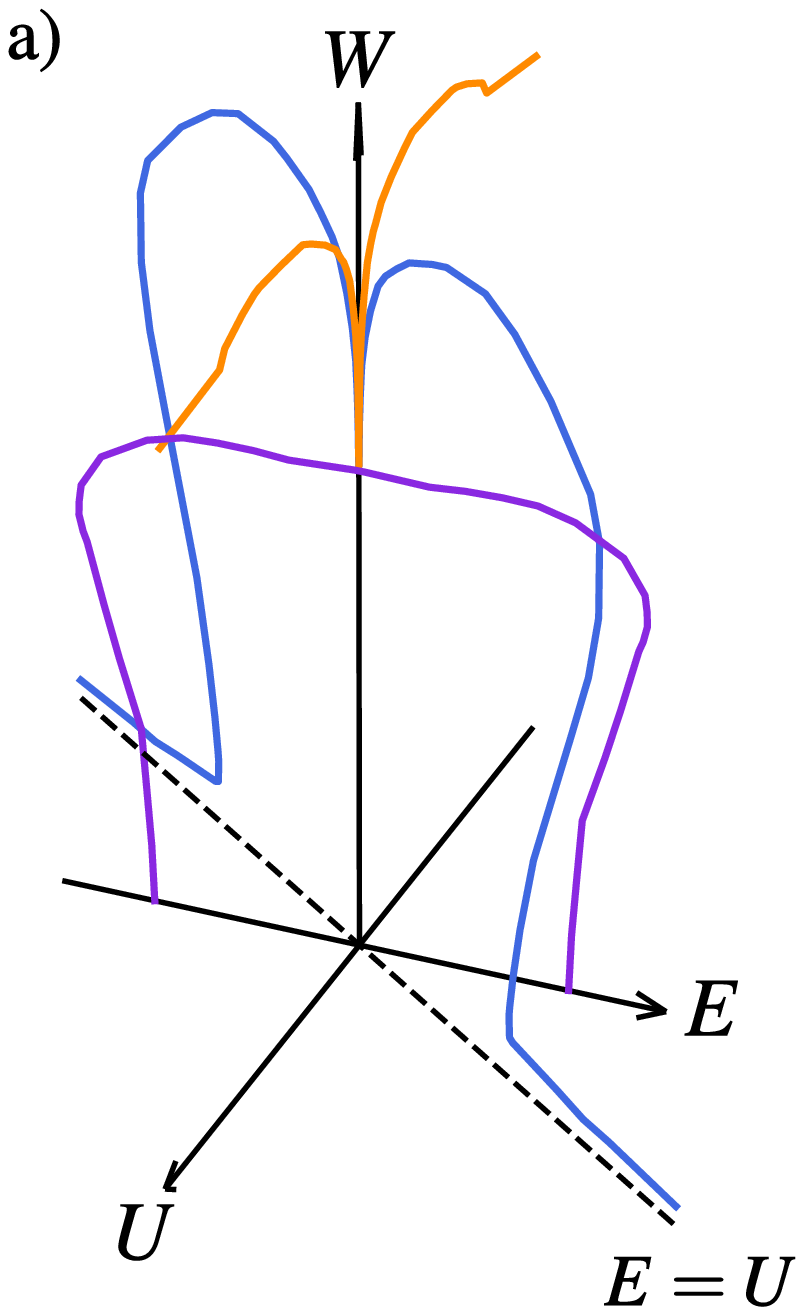}
 	\includegraphics[width=0.625\columnwidth]{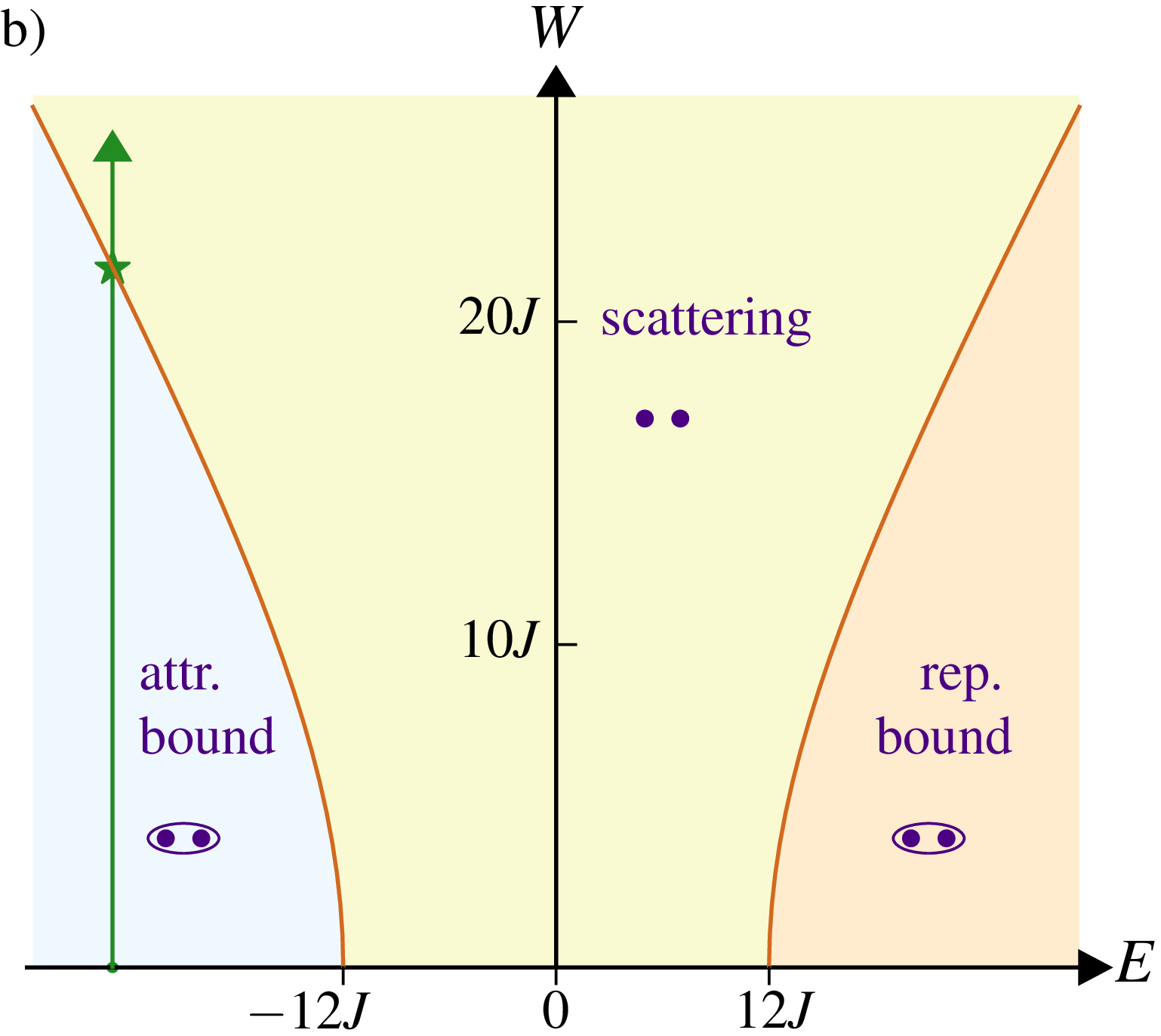}
 	\caption{
 		(a) Critical disorder strength $W_c$ for pair localization as a function of the Hubbard interaction $U$ and the total energy $E$.  The orange and the blue data curves are cuts along the planes $E=0$ and $E=U$. 
 		For vanishing interactions, the phase boundary (violet data curve) coincides with the single-particle 
		mobility edge calculated in Ref.~\cite{Bulka:ZPB1987}, under the change of variable $E=-2\varepsilon$, where $\varepsilon$ is the energy of a single particle.
 		(b) Nature of the pair state as a function of energy and disorder.
 		The two (brown) solid lines define the numerical band edges $E=\pm 2\varepsilon_{\mathrm{be}}(W)$ of the noninteracting two-particle energy spectrum for a given disorder strength $W$. They divide the plane in three regions, corresponding to scattering states, attractively and repulsively bound states. For $|E|>12J$, the nature of the state changes from molecular to scattering-like as the disorder strength increases (solid vertical arrow),  generating multiple Anderson transitions.
 	}
 	\label{Fig:intro}
 \end{figure}

\subsection*{Scope of the paper}
Building on the results of Ref.~\cite{Stellin:PRB2019}, in this work we investigate pairs with nonzero total energy and map out the phase
boundary between localized and extended states in the interaction-energy-disorder space.
This will be done by considering different cuts of the three-dimensional phase diagram along specific planes. 
Some of these cuts are displayed in Fig.~\ref{Fig:intro}(a). 
We see that the critical disorder strength  along the plane $E=U$ (blue line) exhibits an s-like behavior as a function of 
the interaction strength, signaling that in a window of intermediate  $U$ values
 the system undergoes three Anderson transitions as $W$ increases, in contrast with the $E=0$ case (orange line).  

As we shall see, this surprising effect can be explained by the change in the 
nature of the pair state for increasing disorder.  Neglecting  Lifshitz-tail regions, where the single-particle density of states is exponentially suppressed,  the  energy band of a single particle broadens with disorder according to $-\varepsilon_{\mathrm{be}}(W) \leq \varepsilon \leq \varepsilon_{\mathrm{be}}(W)$, 
where the numerical band edges $\pm \varepsilon_{\mathrm{be}}(W)$ are computed for a given disorder strength as explained in Appendix B.
As a consequence, the energy spectrum of \emph{two} noninteracting particles is  bound to the interval $-2\varepsilon_{\mathrm{be}}(W) \leq \varepsilon_1+ \varepsilon_2 \leq 2\varepsilon_{\mathrm{be}}(W)$. 
For given values of $E$ and $W$,  we say that a state is \emph{scattering}-like if the total energy of the pair lies inside 
the two-particle noninteracting spectrum, that is $-2\varepsilon_{\mathrm{be}}(W) \leq E \leq 2\varepsilon_{\mathrm{be}}(W)$. 
These states correspond to the yellow region in the energy-disorder plane shown in Fig.~\ref{Fig:intro}(b).
States which are not scattering-like are called \emph{molecular}. In this case we further distinguish between 
attractively bound states, occurring for $E<-2\varepsilon_{\mathrm{be}}(W)$ and repulsively bound states, which are  defined 
for $E>2\varepsilon_{\mathrm{be}}(W)$; in Fig.~\ref{Fig:intro}b these  states are represented by the cyan and orange regions, respectively.

We see from Fig.~\ref{Fig:intro}(b) that for $|E|<12J$ the pair is described by scattering states for any disorder. The resulting phase diagram at fixed energy is then fairly similar to the $E=0$ case already investigated.
For $|E|>12J$, however, the nature of the pair states changes from molecular  to scattering-like at the disorder threshold $W^*$,  given by the condition $E=\pm 2 \varepsilon_{\mathrm{be}}(W^*)$ (vertical arrow).
We therefore expect Anderson transitions of molecular states at weak disorder, with $W_c<W^*$,  followed by a delocalization transition of scattering states at intermediate disorder, with $W_c>W^*$.
The exploration of these novel metal-insulator transitions of the pair will be the main goal of the present work.

The paper is organized as follows. In Sec. \ref{sec:theory} we review the
underlying theoretical formalism, which amounts to mapping the two-particle Schrodinger equation onto an effective single-particle model  with long-range hopping. 
In Sec. \ref{sec:PD} we present our numerical results for the two-body mobility edge based on transmission-amplitude calculations for elongated bars, while in Sec. \ref{sec:efDOS} we discuss the properties of the disorder-averaged density of states of the effective model.  
Section \ref{sec:conclusion} provides a conclusion and an
outlook.
In Appendix A we present in detail the numerical procedure used to efficiently compute the matrix $K$ of the effective Hamiltonian. In Appendix B we recall the calculation of the  numerical band edge for the (single-particle) Anderson model based on the coherent potential approximation.

\section{Effective single-particle model for the pair}
\label{sec:theory}
 
Hereafter we fix the energy scale by setting $J=1$.
We address the localization properties of the two-body system via a mapping onto an effective single-particle model describing the center-of-mass motion of the pair.
The mapping is exact in the subspace of orbitally symmetric wave-functions, describing either two bosons or two 
fermions in the spin-singlet state (Hubbard interactions have no effect for two fermions in the spin-triplet state).

We start by writing the two-particles Schr{\"o}dinger equation as $(E -\hat H_0)|\psi\rangle=\hat U|\psi\rangle$, where $E$ is the total energy of the pair. 
From Eq.~(\ref{intro1}), we find that the wave-function obeys the following self-consistent equation
  \begin{equation}
\label{formalism2}
|\psi\rangle=\sum_{\mathbf m} U \hat G(E) |{\mathbf m},{\mathbf m}\rangle \langle {\mathbf m},{\mathbf m}|\psi\rangle,  
\end{equation}
where $\hat G(E)=(E \hat I -\hat H_0)^{-1}$ is the noninteracting two-particle Green's function. Equation (\ref{formalism2}) shows that
for contact interactions the  wave function can be completely determined once its diagonal amplitudes
$f_{\mathbf m}=\langle {\mathbf m},{\mathbf m}|\psi\rangle$ are known. By projecting Eq.~(\ref{formalism2}) over the state 
$|{\mathbf n},{\mathbf n}\rangle$, we see that these terms obey a closed equation~\cite{vonOppen:AndLocTIPDeloc:PRL96,Stellin:PRB2019,Dufour:PRL2012,Orso:PRL2005}: 
 \begin{equation}
 \label{integral}
\sum_{\mathbf m} K_{\mathbf n  \mathbf m} f_{\mathbf m} = \frac{1}{U}f_{\mathbf n},
 \end{equation} 
where  $K_{\mathbf n  \mathbf m}  =\langle {\mathbf n},{\mathbf n }|\hat G(E) |{\mathbf m},{\mathbf m}\rangle$. Equation (\ref{integral}) can be interpreted as an \emph{effective}  single-particle  problem with Hamiltonian matrix $K$ and pseudoenergy $\lambda=1/U$, corresponding to the inverse of the interaction strength.
Since $K$ depends explicitly on the total energy, the phase boundary between localized and extended states  of the pair will
represent a surface in the $U-E-W$ space.

The effective model differs from the Anderson model, Eq.~(\ref{Anderson3D}), in two respects. \emph{First}, the matrix elements of $K$ are unknown 
and must be calculated numerically.  When expressed in terms of the eigenbasis of the single-particle model, $\hat H^\textrm{sp} | \phi_r\rangle=\varepsilon_r  | \phi_r\rangle$,
they are given by
\begin{equation}\label{KE0}
K_{\mathbf n  \mathbf m} = \sum_{r,s=1}^N \frac{\phi_{\mathbf n r}  \phi_{\mathbf m r}^*  \phi_{\mathbf n s}  \phi_{\mathbf m s}^*}{E-\varepsilon_r-\varepsilon_s},
\end{equation}
where $\phi_{\mathbf n r} =\langle \mathbf n | \phi_r\rangle$ are the amplitudes  of the wave-functions 
and $N$ is the total number of lattice sites in the grid. \emph{Second}, the matrix $K$ is fully dense, describing hopping processes between arbitrarily distant sites.
 The efficient computation of $K$, which is the main bottleneck
of our approach, is discussed in detail in Appendix A.

For $|E|\gg 1$ or $W\gg 1$, however,  tunneling effects are small and  the effective model  becomes short-range.
To see this, we write the noninteracting two-particle Green's function as
 $\hat{G}(E)=(\hat{A}+\hat{T})^{-1}$, where 
 \be\label{A}
 \hat{A}=\sum_{\textbf{m},\textbf{n}}(E-V_\textbf m -V_\textbf n) |\textbf m, \textbf n \rangle \langle \textbf m, \textbf n |
 \ee
 represents the local part of the Hamiltonian, while
\be\label{T}
  \hat{T} =-\sum_{\textbf{n} \textbf{m}   \boldsymbol \delta} \left (
  \lvert \textbf{n},\textbf{m} \rangle \langle \textbf{n}+  \boldsymbol \delta , \textbf{m} |+   | \textbf{n},\textbf{m} \rangle \langle \textbf{n}, 
  \textbf{m} +\boldsymbol \delta \lvert \right)
 \ee
 accounts for the hopping processes. 
Next, we expand $\hat G(E)$ in powers of $\hat T$, 
retaining up to second orders terms:
 \begin{equation}\label{eqp}
\hat{G}(E) \simeq   \hat{A}^{-1}+ \hat{A}^{-1}\hat{T} \hat{A}^{-1}+  \hat{A}^{-1}\hat{T} \hat{A}^{-1}   \hat{T} \hat{A}^{-1}. 
\end{equation}
The second term in the rhs of Eq.~(\ref{eqp}) does not contribute  to the effective Hamiltonian $K$,
 because $\hat A$ is diagonal in the site basis, whereas 
 $\hat T$ has zero expectation value. 
The third term  contributes through two distinct processes:
(i) a particle hops from a site to a neighboring one and comes back, while the other does not move; 
(ii) both particles move from one site to the same neighboring site, leading to an effective pair hopping. 
An explicit calculation yields 
\begin{eqnarray}
K_{\mathbf n \mathbf m}&&\simeq \frac{1}{E-2V_\mathbf n} \delta_{\mathbf n \mathbf m} +\frac{2}{(E-2V_\mathbf n)(E-2V_\mathbf m)} \times \nonumber \\
&&\sum_{ \boldsymbol \delta} \frac{1}{E-V_\mathbf m-V_{\mathbf m +  \boldsymbol \delta} }
\left(\delta_{\mathbf n \mathbf m}+ \delta_{\mathbf n \mathbf m +  \boldsymbol \delta}\right), \label{atomic}
\end{eqnarray}
where $\delta_{\mathbf n \mathbf m}$ is the Kronecker delta.
The rhs of Eq.~(\ref{atomic}) defines a tight-binding model for the pair, regarded as a point-like particle. In the absence of tunneling,  the matrix $K$ is diagonal, since the two particles can only interact if they share the same lattice site. 

The regime $|E|\gg 1,W$  describes tightly bound states with $E\sim U$. 
 In this limit the  off-diagonal matrix elements in Eq.~(\ref{atomic}) are  approximately constant and equal to $2/E^3$.  
For weak disorder, the effective model (\ref{integral}) reduces to~\cite{Dufour:PRL2012}
 \be\label{Anderson_mol}
\frac{2}{E} \sum_{\boldsymbol \delta } f_{\mathbf n+  \boldsymbol \delta } + \left (2  V_\mathbf n + \frac{4 V_\mathbf n^2}{E}\right ) f_{\mathbf n} \simeq E^2 \left(\lambda-\frac{1}{E}-\frac{12}{E^3}\right) f_{\mathbf n}, 
 \ee
 showing that tightly bound pairs exhibit a quenched tunneling rate,
  $J_b=-2/E$, and feel a  twice larger disorder strength, $W_b=2W$ (neglecting the small 
  $V_{\mathbf n}^2$ correction). Below we will infer the mobility edge of such states from the known~\cite{Bulka:ZPB1987} 
  single-particle results for the 3D Anderson model. 

Equation (\ref{atomic}) applies also to the atomic limit, corresponding to 
 $W\gg 1$. 
In this case the short-range nature of the model is ensured by the fact that the amplitudes 
$\phi_{\mathbf n s} $  of the single-particle wave-functions in Eq.~(\ref{KE0})  have support on very few lattice sites.
Differently from the molecular regime, the pair tunneling rate cannot be seen as approximately uniform,
but
 depends on the specific values of the disorder potential at the two 
edges of the bond. In particular, both diagonal and off-diagonal matrix elements of $K$ can take  large values when the energy denominators in the rhs of Eq.~(\ref{atomic}) become small.

\section{Two-body mobility edge}
\label{sec:PD}

\subsection{Computation of the critical point}

The method followed to extract the position of the mobility edge  has been  presented in detail in Ref.~\cite{Stellin:PRB2019};
here we briefly outline the main steps. 
We consider a bar shaped grid, with  fixed  length $L=150$ and transverse size between $M=8$ and $M=15$, so that $L\gg M$.     
The logarithm of the transmission amplitude, evaluated at a  position $n_z$ along the bar, is defined as~\cite{McKinnonKramer:TransferMatrix:ZPB83}:
\be\label{logT}
F(n_z)=\ln \sum_{\mathbf m_\perp,\mathbf n_\perp} |\langle \mathbf m_\perp,1| G^{\textrm p}(\lambda )| \mathbf n_\perp,n_z \rangle  |^2,
\ee
where $G^{\textrm p}(\lambda)=(\lambda I -K)^{-1}$ is the resolvent of the effective model,
$\mathbf m_\perp =(m_x,m_y) $ and $\mathbf n_\perp= (n_x, n_y)$. We compute the matrix $K$ of the effective Hamiltonian as described 
in Appendix A. 
In order to minimize finite-size effects on the transmission amplitude, the boundary conditions on the single-particle Hamiltonian $H^{sp}$ are chosen periodic in the orthogonal 
directions and open along the transmission axis.
For each disorder realization, we evaluate $F(n_z)$  at regular intervals along the bar and apply  a linear fit to the data,  $f_{\mathrm{fit}}(n_z)=p n_z+q$. 
The Lyapunov exponent is then given by $\gamma_M=-\overline{p}/2$, where $\overline{p}$ is the averaged value of the slope.

The critical point $W=W_c$ of the metal-insulator transition can be identified by  studying the behavior of the reduced localization length $\Lambda_M=1/(\gamma_M M)$
 for increasing values of  the transverse size of the bar. In the metallic phase, $\Lambda_M$ increases as $M$ increases, 
 whereas in the insulating phase it shows an opposite trend.
 At the critical point $\Lambda_M$  converges to a constant  
  $\Lambda_c$  of order unity, depending on the
 universality class and the choice of the boundary conditions. 
In Ref.~\cite{Stellin:PRB2019} we show that our numerical results for $E=0$  are consistent with the orthogonal universality class, where $\Lambda_c=0.576$. This is reasonable, since the effective Hamiltonian $K$ inherits from $H^{sp}$ both the time-reversal and the spin rotational symmetries. 
Finite-size effects, drifting the position of the critical point, are however not negligible in our numerics.  For this reason, 
 the  inclusion of the leading irrelevant variable in the one-parameter scaling ansatz is essential to correctly extrapolate the position of the critical point~\cite{Stellin:PRB2019}.

Below we mainly investigate pair states with total energy $E<-12$.  The case $E>12$ is  recovered  from our study by using the invariance of the Schr{\"o}dinger equation under the transformation $ E \rightarrow -E,  U \rightarrow -U$.
 
 \subsection{Phase diagrams at fixed energy}

 \begin{figure}
\includegraphics[width=\columnwidth]{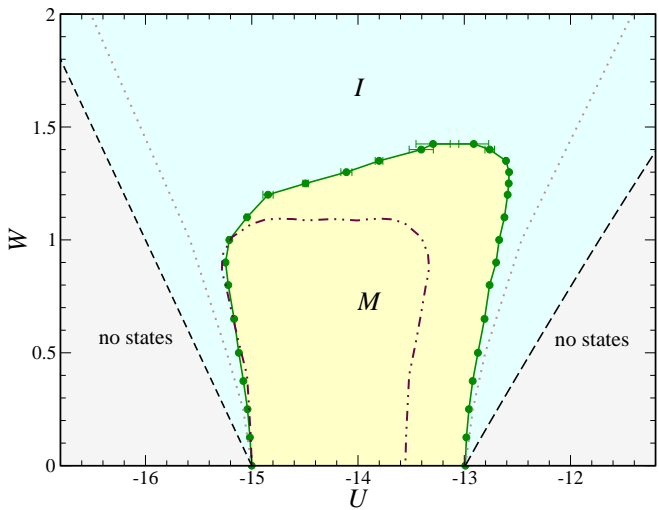}
\caption{Zoom of the interaction-disorder phase diagram for a pair with total energy $E=-15$, in the regime of weak disorder (we set  $J=1$ as energy unit). 
The green circles data refer to the pair mobility edge, separating the metallic (M) phase from the insulating (I) one. The green line is a guide to the eye. The dashed lines correspond to the rigorous edges of the interaction band, $U=E-W$ and $U=1/f(E+W)$, where $f$ is defined in Eq.~(\ref{bound}); below these lines no pair states exist (gray regions).
The dotted lines represent the numerical band edges, which neglects Lifshitz tails, calculated from the disorder-averaged density of states 
of the effective model.
The dot-dashed line corresponds to the mobility edge of pointlike molecules, calculated using the numerical data 
for the single-particle mobility edge from Ref.~\cite{Bulka:ZPB1987}.  
}
\label{Fig:E-15lobe}
\end{figure}

We first present our numerical results for a pair with total energy $E=-15$, focusing initially on the localization properties of the attractively bound states at low disorder. In Fig.~\ref{Fig:E-15lobe} we display the calculated boundary between the metallic (M)  and the insulating (I) phases  (green data points). 
In the absence of disorder, the single-particle wave-functions are plane waves, $\phi_{\mathbf n \mathbf k}=e^{i \mathbf k \mathbf n}/\sqrt{N}$,
with energy dispersion $\varepsilon_{\mathbf k}=-2(\cos k_x+\cos k_y+ \cos k_z)$, where $\mathbf k$ is the lattice momentum. 
From Eq.~(\ref{KE0}) it follows that for $E<-12$ the solutions of the effective Schr{\"o}dinger equation (\ref{integral})  have the same form, $f_{\mathbf n}=e^{i \mathbf Q \mathbf n}$, where
$\mathbf Q$ is the lattice momentum for the center of mass motion. By direct substitution, one finds~\cite{Wouters:PRA2006}
\be\label{RBE}
\lambda=\int \frac{d^3 \mathbf k}{(2\pi)^3} \frac{1}{E-\varepsilon_{\mathbf k} -\varepsilon_{\mathbf Q-\mathbf k} }.
\ee
For $\mathbf Q=0$ and $E<-12$, we can calculate the integral in Eq.~(\ref{RBE}) analytically, by writing the denominator using 
the formula $1/x=-\int_0^{+\infty}e^{xt}dt$, valid for $x<0$. This yields  
$\lambda=f(E)$, where
\begin{eqnarray}
f(E)&=&-\int \frac{d^3 \mathbf k}{(2\pi)^3} \int_0^{+\infty} e^{Et} e^{4 (\cos k_x+\cos k_y+ \cos k_z)t} dt \nonumber \\
&=&-\int_0^{+\infty}e^{Et}\left[\int \frac{d k_x}{2\pi} e^{4 \cos k_x t}\right]^3 dt\nonumber \\
&=& -\int_0^{+\infty}e^{Et}I_0^3(4t) dt\label{bound},
\end{eqnarray}
with $I_n(x)$ being the modified Bessel function of the first kind.  
For $\mathbf Q=(\pi,\pi,\pi)$ the integral in Eq.~(\ref{RBE}) can also be evaluated analytically, because $\varepsilon_{\mathbf k} =-\varepsilon_{\mathbf Q-\mathbf k} $, and therefore
 $\lambda=1/E$. Hence for $W=0$ molecular states exist for $f(E)<\lambda<1/E$, or equivalently, $E<U<1/f(E)$. This is 
 evident in Fig.~\ref{Fig:E-15lobe} by noticing that $1/f(-15)=-12.995$.

The  dashed curves in Fig.~\ref{Fig:E-15lobe} correspond to rigorous  band edges of the system, below which  no states are allowed, due to energy conservation.
To find them, we notice that disorder contributes to the total energy by a term in the interval $[-W,W]$.  Hence the interaction 
band of molecular states for $E<-12$ spreads at most to $E-W<U<1/f(E+W)$. Scattering states are instead possible provided that $-12-W<E<12+W$, independently of the value of the interaction strength. 
By setting $E=-15$, this implies that for $W\geq 3$  all values of the interaction strength are in principle permitted, 
whereas for $W<3$ only states between the two curves  $U=E-W$ and $U=1/f(E+W)$ are allowed.

The two dotted lines in Fig.~\ref{Fig:E-15lobe}  represent the numerical band edge for the pair, calculated from the disorder-averaged 
density of states of the effective model, Eq.~(\ref{integral}).
 The details of the calculation will be presented in Sec.~\ref{sec:efDOS}.
The regions of the phase diagram between
the dotted and the dashed lines correspond to localized states in the  Lifshitz tails regime, where the density of states is very low. 
 
 \begin{figure}
 	\includegraphics[width=\columnwidth]{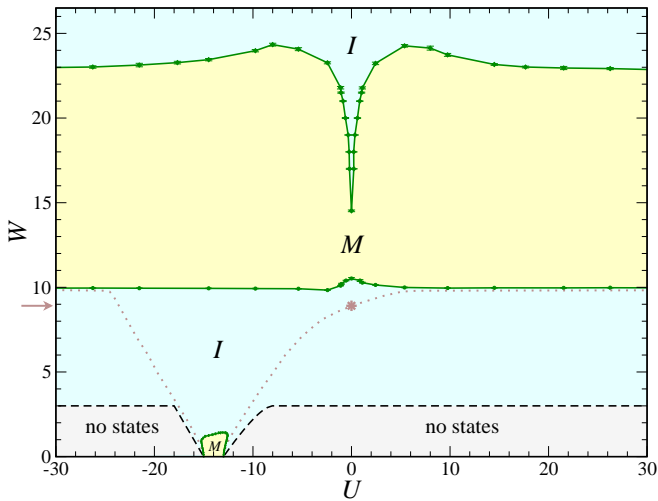}
 	\caption{Complete phase diagram in the interaction-disorder plane for a pair with total energy $E=-15$. The phase boundaries between metallic
 		and insulating phases are displayed by the green symbols. The dashed and dotted lines correspond to the rigorous and the numerical band edges, respectively.
 		The arrow indicates the disorder threshold $W^*=8.91$, where the nature of the pair wave-function changes from molecular to scattering like. At this 
 		point the right numerical band edge  crosses the $U=0$ axis, as indicated by the star symbol.
 		For $3<W<9.8$ (horizontal dashed lines), the pair displays large Lifshitz tail regions. 
 		The remaining notation is the same as in Fig.~\ref{Fig:E-15lobe}. 
 	}
 	\label{Fig:E-15Overall}
 \end{figure}

For comparison, in Fig.~\ref{Fig:E-15lobe}  we also show (dot-dashed line) the prediction for the mobility edge of the pair regarded as a pointlike particle, obeying  Eq.~(\ref{Anderson_mol}).
This is obtained from the numerical data~\cite{Bulka:ZPB1987}  for the 
single-particle phase diagram in the $\varepsilon-W$ plane, taking into account  
 the rescaled energy   $\varepsilon_b=E^2/U-E-12/E$ of the pair as well as  the associated hopping rate $J_b=-2/E$ and disorder strength $W_b=2W$.
The point-like approximation yields very accurate results for pair states near $U=E$, but substantially underestimates the size of the metallic phase for
weaker interactions. 
Indeed, such states describe molecules with lower binding energy, so that the corresponding wave functions can spread over several lattice sites. Figure~\ref{Fig:E-15lobe}  shows that the critical disorder strength for bound states is not
center-symmetric: The tip is shifted towards the right, showing that 
weakly bound pairs are more robust against localization than point-like molecules. 

 We also notice that the point-like approximation misses states at weak interaction, 
 already in the absence of disorder. Indeed, the unperturbed  band edges, obtained from the solution of 
$\varepsilon_b=\pm 6 J_b$, are given by $U=E$ and  $U=E^3/(24 + E^2)=-13.55$ for $E=-15$.
We can  improve the accuracy of the tight-binding model for pairs, by including higher order tunneling terms 
in the rhs of Eq.~(\ref{eqp}). 
The third order term gives zero contribution to the effective Hamiltonian $K$ (like all  odd terms), while the fourth order term gives 
 $\varepsilon_b=E^2/U-E-12/E-240/E^3$ and a larger pair tunneling rate, $J_b=-2/E-120/E^3=0.169$ for $E=-15$. Using this last result,
 the width of the interaction band becomes $12J_b=2.028$, in fairly good agreement with our numerics.
 On the other hand the above fourth order expansion introduces also second-nearest-neighbor hopping processes, which are not contained 
in Eq.~(\ref{Anderson_mol}). These and even longer-range hopping terms become more and more important as the energy $E$ increases 
and the binding energy of the molecule becomes small.

\begin{figure}
\includegraphics[width=\columnwidth]{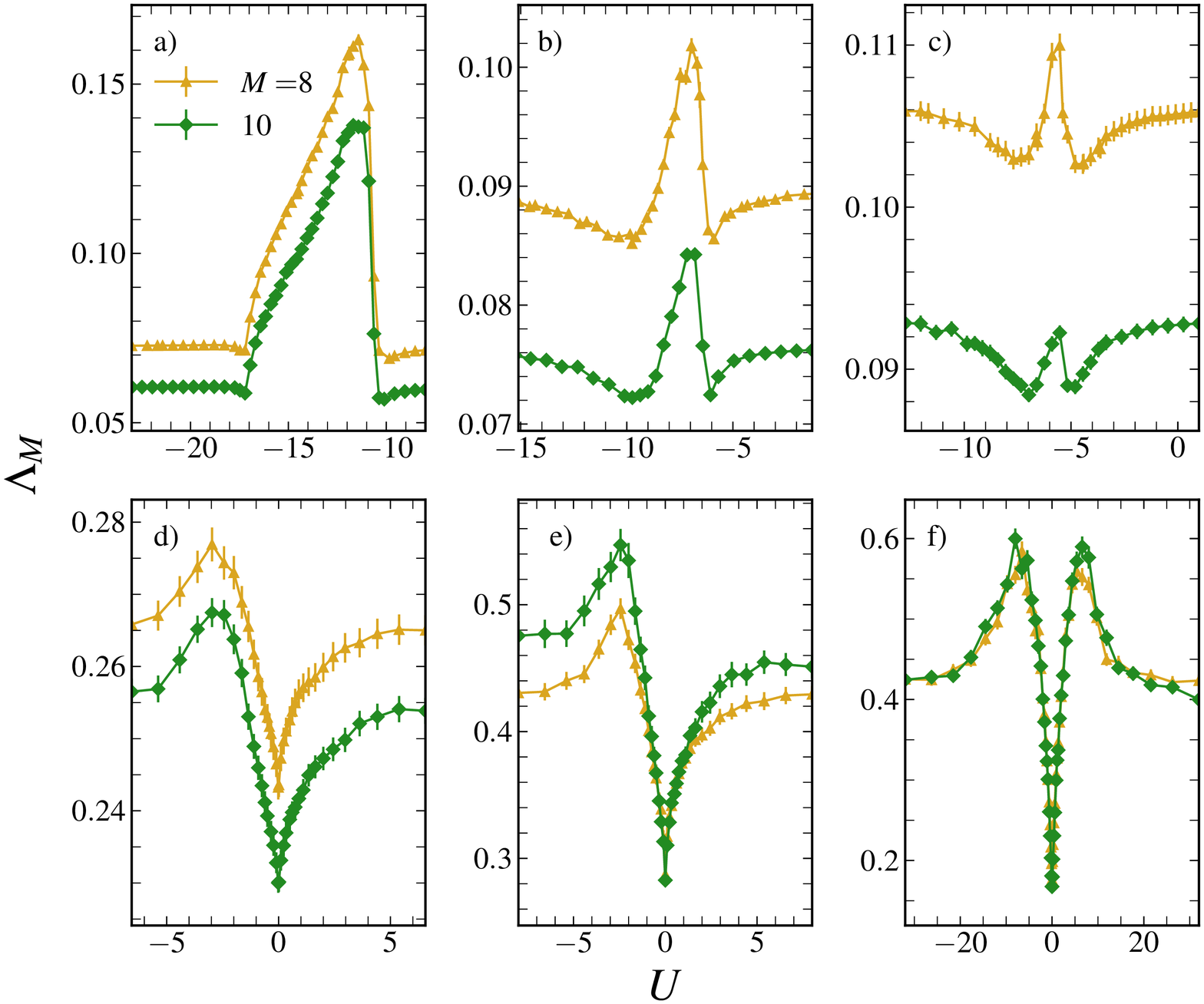}
\caption{Reduced localization length of the pair versus $U$ calculated for two different values of the transverse size of the bar, $M=8$ (circles) 
and $M=10$ (squares). The panels (a)-(f) correspond to increasing values of the disorder strength, $W=3$ (a), $7, 8, 9.4, 10, 24$ (f). }
\label{Fig:LambdaME-15}
\end{figure}

 Let us now discuss the localization properties of the pair  for stronger disorder. The  complete phase diagram for $E=-15$ is shown in
Fig.~\ref{Fig:E-15Overall}.
 In Fig.~\ref{Fig:LambdaME-15} we also display the behavior of the reduced localization length $\Lambda_M$ as a function of the interaction strength, 
 which  helps understanding the structure of the phase diagram.
 The two data curves in each panel correspond to the values $M=8$ and $M=10$ of the transverse size of the bar. The panels (a-f) refer to increasing values of the disorder strength.

We see from Fig.~\ref{Fig:E-15Overall}  that \emph{all} two-particle states are localized for $1.4<W<9.8$. In this insulating phase, the region 
of $U$ values, delimited by the left and right numerical band edges, broadens up as $W$ increases until it covers the entire axis at $W=9.75$. 
 Figures~\ref{Fig:LambdaME-15}(a)-\ref{Fig:LambdaME-15}(c) show that the two curves for $M=8$ and $M=10$ tend to further separate out as $W$ increases, as occurs in the  single particle problem at strong enough disorder (so that asymptotically $\Lambda_M/\Lambda_{M^\prime}=M^\prime/M$). This behavior corresponds to localized molecular states. Interestingly, the same panels show that 
in the Lifshitz tail regions $\Lambda_M$ increases steadily as $W$ increases. 
  
For $W=9.4$   [Fig.~\ref{Fig:LambdaME-15}(d)] the two curves for different $M$ show instead 
 an opposite trend: Their relative distance has reduced,
suggesting that the pair has lost its molecular nature, and is better described by a scattering state. 
  This change of behavior should occur when the energy $E$ of the pair falls inside the noninteracting two-particle energy spectrum, as displayed in Fig.~\ref{Fig:intro}b. The disorder threshold  $W^*$
  is then given by the condition $E=-2\varepsilon_{\mathrm{be}}(W^*)$. We compute the
  single-particle numerical band edge as explained in Appendix B.  The above condition then yields  $W^*=8.91$ for $ E=-15$,  thus confirming the molecule unbinding.
Figure \ref{Fig:E-15Overall} shows that, for $W=W^*$ (horizontal arrow), the 
right numerical band edge for the pair crosses the $U=0$ axis (corresponding to $\lambda\rightarrow \infty$), as indicated 
by the star symbol.

 We see from Fig.~\ref{Fig:LambdaME-15}(d) that at $W=9.4$ the reduced localization length already possesses a clear absolute 
 minimum at $U=0$, which then persists for all larger values of the disorder strength, as displayed in the panels (e) and (f) of the same figure. 
 This confirms that interactions always favor the delocalization of scattering states.  Moreover the delocalization effect is more prominent for intermediate values of the interaction strength, as also occurs in lower dimensional systems~\cite{Frahm1999,Frahm:EigStructAL1DTIP16}.
 By comparing Fig.~\ref{Fig:LambdaME-15}(d) with Fig.~\ref{Fig:LambdaME-15}(e),
 we see that all scattering states are still localized at $W=9.4$, while for $W=10$ they are already all extended, except for few states with 
 vanishing interactions. 
  Figure~\ref{Fig:E-15Overall} shows indeed that the critical disorder strength is nearly constant, $W_c \simeq 9.8$, with a small bump around $U=0$, where $W_c \simeq 10.5$. 
The remarkable overlap between the mobility edge and the numerical band edges 
for strong interactions 
 implies that in this regime pairs possess a large mean free path $\ell$, as follows from the Ioffe-Regel criterion for the metal-insulator transition, $k \ell \sim 1$, $k$ being the (small) effective wave vector of the pair.


The phase boundary at stronger disorder, where the scattering states ultimately localize, is strongly dependent  on the interaction strength, as already observed for the $E=0$ case. In particular states with vanishing interaction are the first to localize around $W\simeq14.5$, 
while for $|U|\gtrsim 2$ the phase transition occurs at much stronger disorder,  between $W= 23$ and $W= 24.5$. Notice that the metallic phase of scattering states is approximately symmetric under the inversion $U\rightarrow -U$. This 
is also clear from Fig.~\ref{Fig:LambdaME-15}(f), showing that the reduced localization length becomes also symmetric under the same transformation.

\begin{figure}
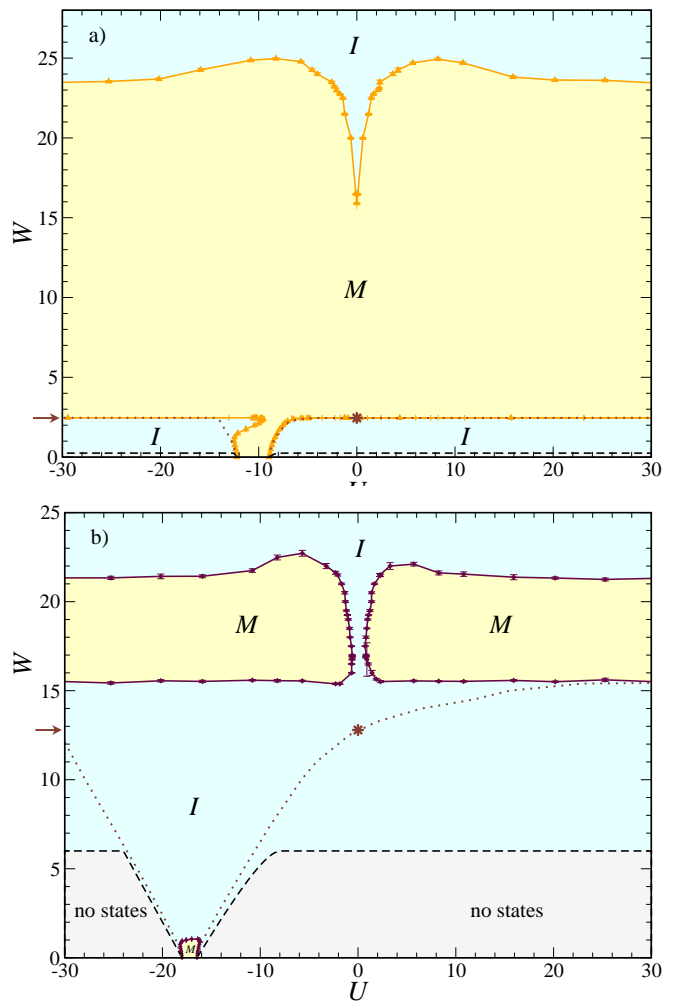

\includegraphics[width=\columnwidth]{msfig5a.eps}
\hspace{0.5cm}
\includegraphics[width=\columnwidth]{msfig5b.eps}
\caption{Topological changes in the phase diagram of the pair for varying energy. Panel (a): phase diagram for $E=-12.25$ showing the two-body mobility edge
(orange up-pointing triangles) together with the rigorous (dashed lines) as well as  the numerical (dotted lines) band edges.  For $U\geq -8.95$ the phase boundary at weak disorder basically superposes with the right numerical band edge. The crossing  from molecular to scattering states occurs at $W=W^*=2.45$, as indicated by the  star symbol. 
Panel (b):  analogous study for $E=-18$. The two-body mobility edge is displayed by the violet diamonds symbols. The disorder threshold for molecular unbinding is $W^*=12.79$. This value is slightly smaller than the prediction $W^*=13.26$
based on the coherent potential approximation, due to finite-size effects.}
\label{Fig:E12m25E18} 
\end{figure}

\begin{figure}
	\includegraphics[width=\columnwidth]{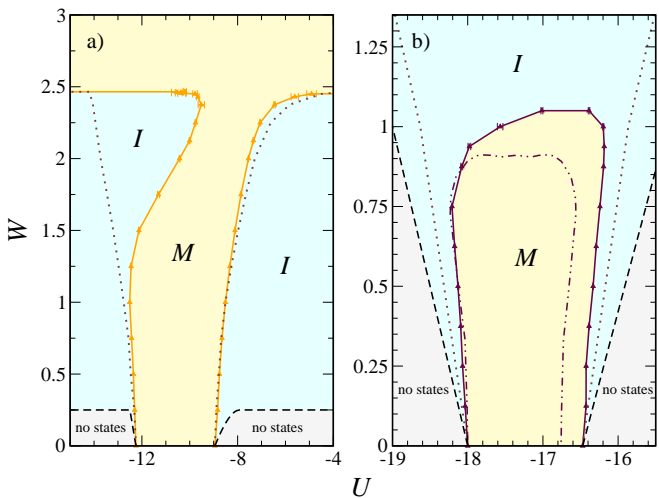}
	\caption{Zoom of the phase diagrams displayed in Fig.~\ref{Fig:E12m25E18} in the low disorder regime.  Panel (a) refers to    $E=-12.25$ and shows that delocalized molecular and scattering states are merged together. Panel	
    (b) displays the molecular mobility edge for $E=-18$ (violet-diamond symbols) together with the prediction based on the 
    point-like approximation for molecules based on Eq.~(\ref{Anderson_mol}) shown by the dot-dashed line. 
	}
	\label{Fig:lobes}
\end{figure}

Let us now explain how the  topology of the phase diagram  in the $U-W$ plane is modified by varying the total energy $E$ of the pair.  
In Fig.~\ref{Fig:E12m25E18}(a) we display the results obtained for $E=-12.25$. In this case the unperturbed band edges are 
given by $U=E$ and $U=1/f(E)=-8.95$.  
A first striking difference with respect to  Fig.~\ref{Fig:E-15Overall} is that the  two metallic phases of molecular and scattering states are merged together.  
Interestingly, for $U\geq -8.95$ the mobility edge at weak disorder closely follows the right numerical band edge. A zoom of the phase diagram in this region is shown in Fig.~\ref{Fig:lobes}(a). We see that localized states which do not belong to Lifshitz tails 
appear only for intermediate values of the disorder strength near the point $U=E$, where the size of the pair is smaller.

We also notice from Fig.~\ref{Fig:E12m25E18}(a)  that  the
unbinding of molecular states and the subsequent delocalization of scattering
states occur almost simultaneously, around $W=W^*= 2.45$.
Hence, for $E\rightarrow -12$, where by definition $W^*=0$, all states at low disorder become extended and the phase diagram becomes qualitatively similar to the 
$E=0$ case, as anticipated in the introduction.
In particular scattering states with vanishing interactions are the first to localize,
 starting at $W \simeq 15.9$.  
A comparison with Fig.~\ref{Fig:E-15Overall} reveals that the maximum value of the associated critical disorder strength 
shifts to  weaker interactions, as the energy $E$ decreases.

Next, we explore the shape of the phase diagram in the opposite limit, where the energy  of the pair is instead large and negative.
In Fig.~\ref{Fig:E12m25E18}(b) we show the obtained results for $E=-18$. 
In this case the metallic phase of scattering states splits out in two disconnected parts, with support at positive and negative $U$ values, respectively, implying that there are no metallic pair states for vanishing interactions. 
These regions of delocalized scattering states shrink rapidly in size as $|E|$ increases. For instance  we see from Fig.~\ref{Fig:E12m25E18}(b) that pair states become scattering like at $W=W^*=12.79$, while the delocalization transition  occurs only for $W \gtrsim 15.5$; in contrast, such states are more easily localized at stronger disorder, the last Anderson transitions occurring around $W=22.5$.
By comparing Fig.~\ref{Fig:E12m25E18}(b) with Fig.~\ref{Fig:E-15Overall} and  Fig.~\ref{Fig:E12m25E18}(a), we also notice that the phase boundary of scattering states is also less smooth. This is due to the fact that, when both $|E|$ and $W$ take large values, residual finite-size effects, which are not completely removed by the scaling procedure, start to appear. We attribute this behavior to the fact that in this regime only few strongly localized single-particle states contribute significantly to the kernel $K$ in Eq.(\ref{KE0}), by making the energy denominator small. As a consequence, the
reduced localization length exhibits larger statistical error bars, and so does the position of the critical point, obtained from the finite-size scaling procedure.

In Fig.~\ref{Fig:lobes}(b) we display the molecular band edge for $E=-18$ (violet-diamond symbols) together with
the prediction based on the point-like approximation for molecules (dot-dashed line). As compared to the case $E=-15$, shown
in Fig.~\ref{Fig:E-15lobe}, the two-body mobility edge is more center-symmetric  and the 
point-like prediction works considerably better. 

\subsection{Phase diagram along the $E=U$ plane} 
\begin{figure}
\includegraphics[width=\columnwidth]{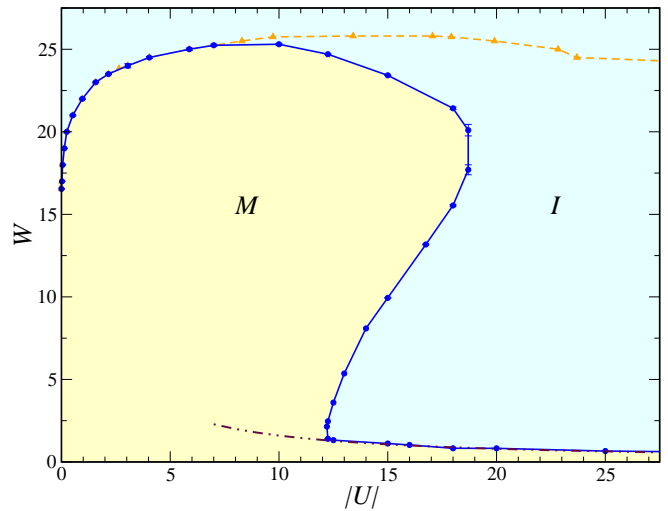}
\caption{Phase diagram in the interaction-disorder plane for a pair with total energy $E=U$ (blue circles data). 
The orange triangles data refer to the phase boundary at $E=0$, calculated in Ref.\cite{Stellin:PRB2019}.
The double dot-dashed  line at low disorder corresponds to the molecular result, $W_c\simeq 16.0/|U|$, obtained by treating the pair as a 
point-like particle obeying an effective Anderson model, see Eq.~(\ref{Anderson_mol}). The diagram holds for both attractive and repulsive interactions.
}
\label{Fig:E=U}
\end{figure}

We now proceed to discuss the cut of the three-dimensional phase diagram of the pair along the $E=U$ plane, which was anticipated in Fig.~\ref{Fig:intro}(a).  The
same numerical data are displayed in Fig.~\ref{Fig:E=U} (blue circles) together with the previous results for $E=0$ (up orange triangles).
 While for weak interactions the two data curves remain very close,  their behavior in the strongly interacting regime is completely different. For $E=U$ we see that the phase boundary displays a double reentrant (s-like) behavior in the interval $12<|U|\lesssim 19$ (we recall that the diagram is symmetric 
under $U\rightarrow -U$). 
Here the two-particle system undergoes three metal insulator transitions
as the disorder strength increases, corresponding to localization of molecules,
delocalization and subsequent localization of scattering states, respectively. These critical points are obtained from Fig.~\ref{Fig:E-15Overall} and Figs.~\ref{Fig:E12m25E18}(a) and
~\ref{Fig:E12m25E18}(b) by intersecting the phase boundary with the vertical line at $U=E$.

It is interesting to note that the critical disorder strength for the localization of molecules with $E=U$ can be easily computed from the point-like approximation based on Eq.~(\ref{Anderson_mol}). Indeed, from the data of Ref.~\cite{Bulka:ZPB1987} the critical disorder strength at the unperturbed left band edge is $W_c^{sp}(\varepsilon=-6)\simeq 16$. By expressing it in terms of the molecular parameters, we obtain $W_c \simeq 16/|U|$. 
This is shown in Fig.~\ref{Fig:E=U} by the violet double dot-dashed line, which is in very good agreement with our numerics for $|U|>12$.


\subsection{Recovering the single-particle mobility edge}

\begin{figure}
	\includegraphics[width=\columnwidth]{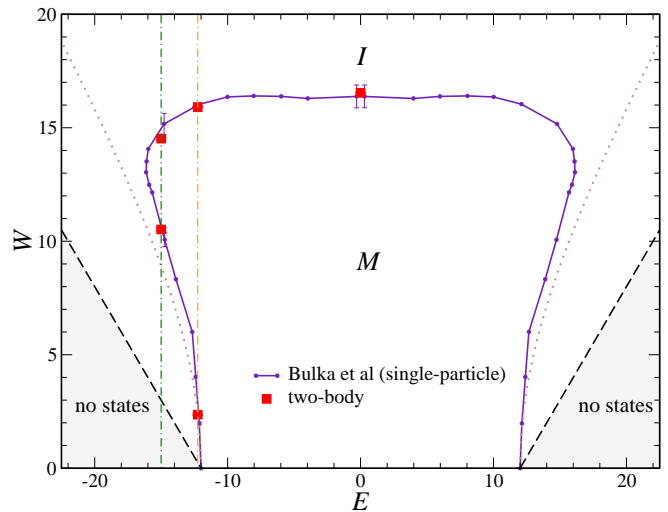}
	\caption{Comparison between two-body and single-particle mobility edges for vanishing interactions. The red square symbols denote the 
	two-body data calculated for total energies $E=-12.25$ (orange dashed line) and $E=-15$ (green dashed line).  For $E=-18$ no critical point is found. The result  for $E=0$ obtained in Ref.\cite{Stellin:PRB2019} is also shown.      The continuous violet line is a guide to the eye connecting  the numerical data for the single-particle phase boundary  
	extracted from Ref.~\cite{Bulka:ZPB1987}, upon the change of variable $E=2\varepsilon$, $\varepsilon$ being the single-particle 
	energy. The dashed lines correspond to the rigorous band edges $W=-12\pm E$, while the dotted lines refer to the numerical band edges of the pair for $U=0$ [displayed as solid lines in Fig.~\ref{Fig:intro}b]. 
}
	\label{Fig:U=0}
\end{figure}

A natural question that arises from our discussion is: How does the two-body phase diagram in the $E-W$ plane behave in the limit of vanishing interactions? What is the explicit connection with the single-particle mobility edge in the $\varepsilon-W$ plane? 
The answer to this question is shown in Fig.~\ref{Fig:U=0}, where the data symbols correspond to the critical points at vanishing interactions obtained 
for $E=-15$ and $E=-12.25$ (vertical dashed lines) 
from the numerical data of Fig.~\ref{Fig:E-15Overall} and Fig.~\ref{Fig:E12m25E18}(a) (we recall that for $E=-18$ there are no transitions as $U\rightarrow 0$). The corresponding result for  $E=0$ has also been added.
The continuous violet line in Fig.~\ref{Fig:U=0} is a guide to the eye of the numerical data for the single-particle mobility edge obtained in Ref.~\cite{Bulka:ZPB1987}, expressed in terms of the pair energy $E=2\varepsilon$.  
We see that for vanishing interactions, our numerical results for the two-particle mobility edge  are fully consistent (within the numerical accuracy) with the single-particle counterpart.  

Our results for the rigorous and the numerical band edges also agree  with the single-particle picture. For instance, the rigorous band edges of the pair for $U\rightarrow 0$ are given by the equations $-12-W\leq E\leq 12+W$, which is equivalent to  $-6-W/2\leq \varepsilon \leq 6+W/2$.
The numerical band edge at $W=W^*$, corresponding to the crossing from molecular to scattering states,  is fixed by the condition  $E=\pm 2\varepsilon_{\mathrm{be}}(W)$, yielding $\varepsilon=\pm \varepsilon_{\mathrm{be}}(W)$, as expected.

\section{Density of states of the effective model}
\label{sec:efDOS}

The disorder-averaged density of states (DOS) of the effective model for the pair, expressed as a function of the inverse interaction strength $\lambda=1/U$, is defined as
\be\label{DOS}
\rho_K(\lambda)=\frac{1}{N} \sum_{r=1}^N  \overline {\delta(\lambda-\lambda_r)}, 
\ee
where  $\lambda_r$ are the eigenvalues of the kernel $K$ for a given disorder realization and the bar indicates the average over the different disorder realizations.
Although this quantity  does not show any singular behavior at  the critical point of the Anderson transition, it provides useful information on the distribution of the (pseudo)energy levels which can help us understanding the 
two-particle phase diagram.
While the computation of the transmission amplitude  requires bar-shaped grids,   the DOS can be calculated more accurately using cubic lattices, with $L=M$,  
assuming periodic boundary conditions along the three directions. To this end, we compute the matrix $K$ of the effective model with the help of the
Woodbury matrix identity, as  discussed in Appendix A. 

We evaluate the DOS numerically by partitioning the interval $[\lambda_\textrm{min},\lambda_\textrm{max}]$, where it is significantly 
different from zero, into $N_b$ bins of equal width $\Delta\lambda=|\lambda_\textrm{max}-\lambda_\textrm{min}|/N_b$.  The number of bins used
 for the evaluation is chosen of the order of the square root of the number of data points per disorder realization,  $N_b\sim \sqrt{N}$. 
 Let $\lambda_j=\lambda_\textrm{min}+\Delta\lambda (j-1)$ label the points of the grid, with $j=1, ..,N_b$ and let $N_{tr}$
be the total number of disorder realizations considered (in our case $N_{tr}=200$). For each bin $j$ and for each disorder realization $r$, with $r=1,..,N_{tr}$, we count the relative number of occurrences $p_j^{{r}}$, corresponding to the ratio
 between the number of eigenvalues of the matrix $K$ falling inside the bin and the total number $N$ of eigenvalues. 
 The corresponding value of the DOS  is calculated as
 \be
 \rho_K (\lambda=\lambda_j)=\frac{1}{N_{tr}\Delta\lambda} \sum_{r=1}^{N_{tr}} p_j^{{r}},
 \ee
where the factor $\Delta\lambda$ in the rhs ensures the correct normalization condition,  $\int_{-\infty}^{+\infty}\rho_K(\lambda) d\lambda = 1$.

 \begin{figure}
 	\includegraphics[width=\columnwidth]{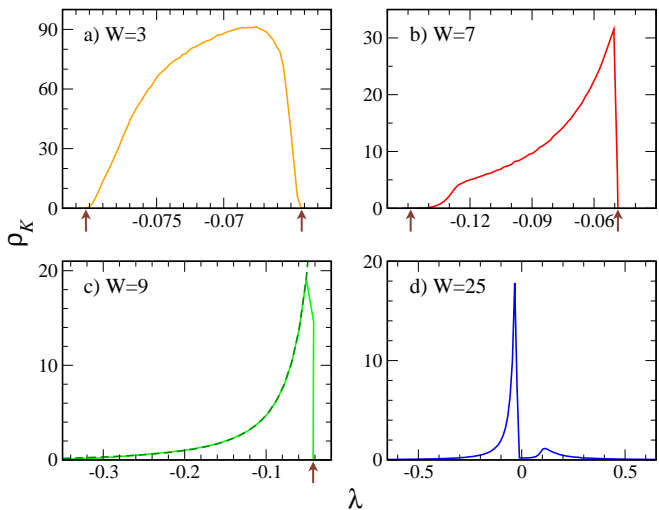}
 	\caption{Disorder-averaged density of states $\rho_K$ of the effective Hamiltonian for the pair, see Eq.~(\ref{DOS}), as a function 
 		of $\lambda=1/U$. The four panels correspond to increasing values of the disorder strength, while the total energy is fixed to $E=-15$. The calculation is done assuming a cubic box of sizes $L=M=24$ with periodic boundary conditions. The vertical arrows indicate the positions of the numerical band edges, where the Lifshitz tails regions appear. The dashed line in panel (c) corresponds to a power-law fit of the left tail of the data with $\rho_K^\textrm{fit}(\lambda)=a_0\lambda^{a_1}$ yielding $a_0=0.045\pm0.02$ and $a_1=-2.03\pm0.14$.}
 	\label{Fig:rhoE-15}
 \end{figure}

In Fig.~\ref{Fig:rhoE-15} we display  the DOS of a pair with total energy $E=-15$ for increasing values of the disorder strength [panels (a)-(d)]. The vertical arrows  mark the position of the numerical band edges, signaling the crossing to a Lifshitz tail region.
In this work we assume that a given bin $j$ belongs to the Lifshitz tails region if the corresponding value of the DOS satisfies
\be\label{tail}
\rho_K(\lambda_j) < \frac{C}{\Delta\lambda N},
\ee
where $C$ is a constant of order unity, which for definiteness we choose equal to $C=1/2$. 
The numerical band edges are then obtained as the borders of the region of the $\lambda$ spectrum, where  Eq.~(\ref{tail}) is satisfied.
We have checked that, for the single-particle Anderson model, this working procedure yields results which are consistent with the prediction based on the coherent potential approximation~\cite{Kroha:SelfConsistentTheoryAnderson:PRB90}.

\begin{figure}
	\includegraphics[width=\columnwidth]{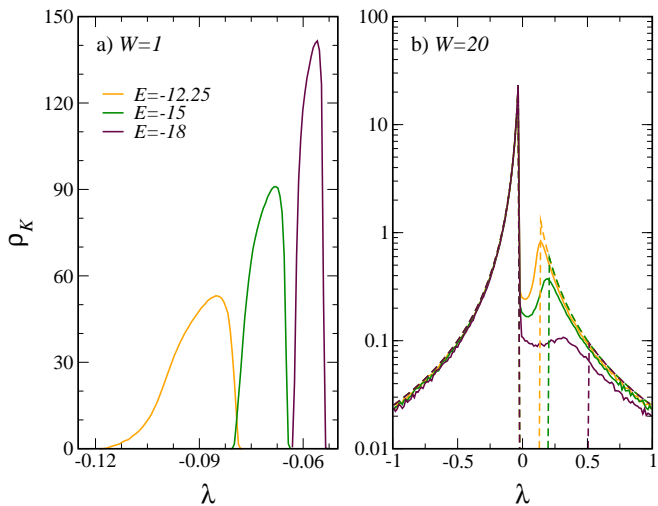}
	\caption{Disorder-averaged density of states of the effective Hamiltonian for the pair, see Eq.~(\ref{DOS}), as a function 
		of $\lambda=1/U$, calculated for three different values of its 
		total energy $E=-12.25$ (orange line), $-15$ (green line), and $-18$ (violet line). The left panel (a) corresponds to $W=1$, while the right panel (b) refers to $W=20$.
		The grid used is the same as in Fig.~\ref{Fig:rhoE-15}.}
	\label{Fig:rhoLambdaEvar}
\end{figure}

For very weak disorder [panel (a)],  the DOS is nonzero only in a narrow region  around $\lambda=1/E=0.0667$, as expected for a tightly bound state. 
For fixed $W$, the DOS broadens as the modulus  $|E|$  of the energy diminishes, because molecules are less bound, as shown in Fig.~\ref{Fig:rhoLambdaEvar}(a)
for $W=1$.
The DOS also broadens as the disorder becomes stronger. This effect  is clearly visible in the phase diagram of Fig.~\ref{Fig:E-15Overall}, where the dotted lines represent the numerical band edges expressed in terms of the interaction strength $U=1/\lambda$. 
For instance, for $E=-15$ and $W=7$, we see from Fig.~\ref{Fig:rhoE-15}(b) that the Lifshitz tails region is given by $\lambda <-0.149$ and $\lambda >-0.0485$, which translates to $-20.64 < U < -6.72$.

As molecules turn into scattering states, at $W=W^*=8.91$, the support of the DOS becomes unbound, due to the presence of a long-range tail, as shown in Fig.~\ref{Fig:rhoE-15}(c). A power law fit to the tail reveals that the DOS  decays algebraically as 
$ \lambda^{-2}$, as  displayed in the same panel with the dashed line.
This asymptotic behavior signals that the DOS, expressed 
in terms of the interaction strength as $\tilde \rho_K(U)=\rho_K(\lambda)\lambda^2$, becomes non zero in the noninteracting limit, $\tilde \rho_K(0)\neq 0$; it is therefore a specific 
feature of the scattering nature of the pair.

For stronger disorder, states for repulsive interactions $(\lambda>0)$ become also available, as shown in Fig.~\ref{Fig:rhoE-15}(d) for $W=25$. Differently from the behavior of the reduced localization length (see Fig.~\ref{Fig:LambdaME-15}), the DOS remains strongly asymmetric under a parity transformation $\lambda\rightarrow -\lambda$, even for rather large values of the disorder strength.
This feature can be better understood starting from the atomic limit,
where tunneling terms in Eq.~(\ref{atomic}) can be neglected, so that 
the matrix $K$ becomes diagonal and the DOS can be computed analytically~\cite{Stellin:PRB2019}
%
 \begin{equation} \label{rhoSR}
 \rho_K(\lambda)= \frac{1}{2W\lambda^2}\Theta\left(W-\left|E-\frac{1}{\lambda}\right|\right),
 \end{equation} 
where $\Theta$ is the  unit step function. Equation~(\ref{rhoSR}) confirms that the DOS behaves as $\lambda^{-2}$, but states with small 
$\lambda$ are forbidden due to the energy conservation, $|E-1/\lambda|< W$.

An explicit comparison of Eq.~(\ref{rhoSR}) with the full numerical computation of the DOS is shown in Fig.~\ref{Fig:rhoLambdaEvar}(b) for 
$W=20$ and for three different values of the total energy $E$ of the pair. 
The vertical dotted lines refer to the support of the DOS obtained from
Eq.~(\ref{rhoSR}). We see that, for almost all negative values of $\lambda$, the DOS is essentially independent of the energy, as expected. The agreement is less good
in the strongly interacting regime, corresponding to vanishing $\lambda$.  
Here tunneling effects are important and lead to a finite value of the DOS, $\rho_K(0)>0$. In contrast, the power-law  tails are rather insensitive to such  effects, since hopping can always be regarded as perturbative for $\lambda\rightarrow\infty$. 
From Fig.~\ref{Fig:rhoLambdaEvar}(b) we further notice that  the DOS becomes more symmetric as the modulus  $|E|$ of the total energy decreases. A full symmetry, however, is recovered only for $E=0$~\cite{Stellin:PRB2019}.

 \section{CONCLUSION AND OUTLOOK}
 \label{sec:conclusion}
In this work we have investigated the localization properties of  two identical bosons or two fermions with opposite spins moving in a disordered three-dimensional lattice and subject to onsite interactions. 
The two-body Anderson-Hubbard model provides the simplest example of Anderson transitions in three-dimensional interacting quantum systems. 
Our theoretical approach is based on an exact mapping of the original Hamiltonian into an effective single-particle model with long-range hopping, describing the center-of-mass motion of the pair. The  critical properties of the effective model are investigated numerically via large-scale simulations (approximately 1.5 million hours of CPU time in state-of-the-art supercomputers).

We found that the two-particle phase diagram in the interaction-energy-disorder space presents an incredibly rich structure characterized by multiple metallic and insulating phases.  We showed that this effect originates from the change in the nature of pair states, from molecular to scattering-like, as the disorder strength increases.
%
Our work provides a general framework to study the mobility edge of molecules of arbitrary size, going beyond the point-like approximation holding in the strongly interacting regime. In particular, it allows us to  describe  the behavior of the pair near the dissociation threshold and its subsequent  delocalization as a scattering state.  

Some of our results can readily be  tested in current experiments~\cite{Kondov:PRL2015} simulating the three-dimensional fermionic Anderson-Hubbard model with atomic gases, by using ultradiluite samples. These include the observation of interaction-induced delocalization of pairs in regimes where all single-particle states are localized as well as the localization of either attractively or repulsively bound states at low disorder.  

We hope that our work will contribute to bridge together the field of few-body Anderson localization with its many-body counterpart, at finite particle density. In particular, if  a many-body mobility edge exists for the three-dimensional Anderson-Hubbard model, its behavior in the zero-density limit must be consistent with the predictions of few-body physics. Notice that the two-body mobility edge discussed here appears only in the subspace of orbitally symmetric two-particle wave functions, describing either bosons or fermions in spin-singlet state; here interactions can induce a delocalization transition of the system even if all single-particle states are localized. In contrast, fermions in spin triplet states localize  as noninteracting particles. 
We also point out that the localization properties of the pair were inferred from the behavior of the diagonal 
amplitudes $\langle {\mathbf m},{\mathbf m}|\psi\rangle$ of the wave-function. 
Recently, it has been shown~\cite{krause2019nucleation}  that, for sufficiently low disorder, 
a single spin-down fermion is sufficient to thermalize a one-dimensional localized bath of spin-up fermions, through the propagation of the doublon excitation; a similar effect was also shown to apply for bosonic systems. It would be interesting to study (both numerically and experimentally)
the same mechanism in three dimensions,  and obtain the  many-body mobility edge as a function of the bath density. 
Our two-body prediction will then be recovered in the limit of vanishing bath density.



In this work we have considered the case of contact interactions, Eq.~(\ref{intro1}). The effective model could be generalized to include 
nonlocal interactions, for instance between neighboring sites, provided the interaction Hamiltonian  can still be written as
$\hat U =U \hat P$, where $\hat P$ is a projector operator, as considered in Ref.~\cite{Frahm:EigStructAL1DTIP16}.
Finally, our approach can be adapted to investigate the transport properties of other kinds of two-particle systems subject to quenched randomness, like Cooper pairs in strongly disordered atomic gases~\cite{Cao:PRA2016} or  superconductors~\cite{PhysRevB.62.8665,Feigelman:PRL2007,Sacepe:NatPhys2011}.
 Investigations of the steady-state and out-of-equilibrium properties of a Fermi gas undergoing the BCS-BEC crossover in the presence of a random potential~\cite{PhysRevLett.99.250402} are already under way~\cite{Krinner:PRL2015,Nagler:NJP2020,nagler2019dipole,nagler2019unraveling}.

%
%



\section*{ACKNOWLEDGEMENTS} 
We acknowledge D. Delande, K. Frahm, C. Monthus, S. Skipetrov and T. Roscilde for fruitful discussions.
This project has received funding from the European Union's Horizon 2020 research and innovation programme under the 
Marie Sklodowska-Curie Grant agreement No. 665850. This work was granted access to the HPC resources of CINES (Centre Informatique National de l'Enseignement Sup\' erieur) under the allocations 2018-A0040507629, 2019-A0060507629, and 2020-A0080507629 supplied by GENCI (Grand Equipement National de Calcul Intensif).

\section*{Appendix A: NUMERICAL evaluation of the matrix $K$}
\label{sec:numerics}

In this subsection we outline the numerical procedure followed to efficiently compute the entries of the effective Hamiltonian matrix $K$ for the pair. 
We consider a grid of length $L$ and squared transverse section of length $M$, with $L\leq M$.
We evaluate the effective Hamiltonian from Eq.~(\ref{KE}), by writing the matrix elements as~\cite{Frahm1999}
\begin{equation}\label{KE}
K_{\mathbf n  \mathbf m} = \sum_{r=1}^N \phi_{\mathbf n r}  \phi_{\mathbf m r}^*  \langle \mathbf n | G^\textrm{sp}\mathbf (E-\varepsilon_r) | {\mathbf m}\rangle,
\end{equation}
where $G^\textrm{sp}(\varepsilon)=(\varepsilon I_d -H^\textrm{sp})^{-1}$ is the resolvent of the Anderson model, and $I_d$ is the identity matrix.
 Equation~(\ref{KE}) shows that the evaluation of the effective Hamiltonian $K$ requires $N$ inversions of $N\times N$ matrices, implying that the computational complexity 
 is $O(N^4)$.  
  Fortunately, we can accelerate the calculation of the resolvent exploiting specific properties of the single-particle Hamiltonian, $H^{sp}$.
 In the presence of open boundary conditions along the longitudinal direction, the latter possesses a block-tridiagonal structure, each block corresponding to a transverse section of the bar.  
As a consequence, the resolvent can be written as  
   \be
G^\textrm{sp}= {\begin{pmatrix}
    A_{1} & 1 & 0 & \dots  & 0 & 0 \\
    1 & A_2 &  1 & \dots  & 0  & 0 \\
    0 & 1 & A_3 &  \dots  & 0  & 0 \\
    \vdots & \vdots & \vdots & \ddots & \vdots& \vdots \\
    0 & 0  &0 & \dots &  A_{L-1} &1 \\
    0 & 0 & 0 &  \dots & 1  & A_L
\end{pmatrix}}^{-1} \label{Gtri},
 \ee
 where $A_i$ are $M^2\times M^2$ symmetric matrices defined by
 \be
 A_i=\varepsilon I- H^{sp}_i,
 \ee
with $H^{sp}_i$ being the the Hamiltonian matrix of the $i-$th block, while $1$ and $0$ are the identity and the zero matrices, respectively. 

Matrices as in Eq.~(\ref{Gtri}) can be efficiently inverted using recursive techniques~\cite{Jain:2007}. To do so, we first compute a sequence of symmetric matrices $S_i$, with $i=1, .., L-1$, using the recurrence relation
\be
S_i=(A_{i+1}-S_{i+1})^{-1},
\ee
starting from $S_{L-1}=A^{-1}_L$. Let $D_i$ and $C_{ij}$ be, respectively, the  diagonal and off-diagonal blocks of the matrix $G^\textrm{sp}$
that we want to compute (since $G^\textrm{sp}$  is symmetric, we can restrict to $i>j$). These matrices can be determined using 
the coupled recursive relations
\begin{equation}
\nonumber
\begin{pmatrix} C_{i1} & C_{i2}&\dots & C_{i i-1}\\\end{pmatrix}=-S_{i-1}\begin{pmatrix}C_{i-11}&C_{i-12}&\dots &D_{i-1}\\ \end{pmatrix}
\end{equation}
and 
\begin{equation}
\label{eqn:2011}
D_{i+1}=S_i (1+D_i S_i),
\end{equation}
for $i=2, .., L-1$, starting from $D_1=A_1-S_1$.
Using the above procedure, the computational complexity to find $G^{sp}$ reduces 
to $O(L^2M^6)$, so that  the overall cost to evaluate the full kernel $K$ scales with the system size as $L^3M^8$.  

Let us now consider the case of periodic boundary conditions along the longitudinal direction. In this case 
the matrix to invert differs from the rhs of Eq.~(\ref{Gtri}) by two non vanishing block entries, ${G^{sp}}^{-1}_{1L}={G^{sp}}^{-1}_{L1}=1$. 
 Although such a matrix  is no longer block-tridiagonal, it can still be inverted efficiently. To see this, we write it as 
 ${G^{sp}}^{-1}=(B +U^t V)^{-1}$, 
where $B$ is a block-tridiagonal matrix obtained from the rhs of Eq.~(\ref{Gtri}) under the change  $A^\prime_1=A_1+A_L^{-1}$ and $A^\prime_L=2A_L$, while
\begin{eqnarray}\nonumber 
U&=& \begin{pmatrix} -A_L^{-1} &0 & \dots &0& 1 \end{pmatrix} \\
V&=& \begin{pmatrix} 1 & 0 & \dots  &0 &-A_L \end{pmatrix} 
\end{eqnarray}
are $M^2\times N$ matrices. After computing the inverse of $B$ using the above procedure, 
 we determine the resolvent $G^{sp}$ via the  Woodbury matrix identity:
\be
(B +U^t V)^{-1}=B^{-1}-B^{-1}U^t (1+V B^{-1}U^t)V  B^{-1}. \label{woodbury}
\ee
The second term in the rhs of Eq.~(\ref{woodbury}) can be calculated using $M^2N^2$ elementary operations, which corresponds to the
 same computational complexity  $O(M^6L^2)$ of inverting $B$.  This is consistent with our numerical experiments showing that, in the presence 
 of periodic boundary conditions along the bar, the time needed to evaluate the effective Hamiltonian approximately doubles.

\section*{Appendix B: single-particle numerical band edge} 

Neglecting  Lifshitz tails, the numerical band edge $\varepsilon_{\mathrm{be}}(W)$ 
for the Anderson model, Eq.~(\ref{Anderson3D}), can be accurately estimated via the coherent potential approximation (CPA) as done in Ref.~\cite{Kroha:SelfConsistentTheoryAnderson:PRB90}. Here we review the main steps for completeness.
We begin by expressing the diagonal term $\mathcal{G}(\varepsilon)=\overline {\langle \mathbf n | (\varepsilon I -\hat H^\textrm{sp})^{-1}   |\mathbf n\rangle }$ 
of the disorder-averaged (translationally invariant) single-particle Green's function as 
$\mathcal{G}(\varepsilon)=\mathcal{G}_0(\varepsilon-\Sigma)$, where 
\be\label{G0}
\mathcal{G}_0(\varepsilon)=\int_{-\pi}^\pi \frac{d^3k}{(2\pi)^3} \frac{1}{\varepsilon-\epsilon(\mathbf k)+i 0}
\ee
is the disorder-free counterpart and $\Sigma$  is the self-energy.
The latter can be found by solving the (self-consistent) CPA equation
 \be\label{CPA0}
 \int dV P(V) \frac{1}{1-(V-\Sigma)\mathcal{G}}=1.
 \ee

By substituting the box random potential distribution (\ref{randombox}) in Eq.~(\ref{CPA0}) and performing the integration 
over the disorder amplitude, we end up with the following equation
 \be
 \label{CPA}
 \ln \left(\frac{2-\mathcal{G} W+2 \mathcal{G} \Sigma}{2+\mathcal{G} W+2 \mathcal{G}  \Sigma} \right)+   \mathcal{G} W=0,
  \ee
 whose solution yields the self-energy as a function of the single-particle energy and the disorder strength, $\Sigma=\Sigma(\varepsilon,W)$.
The multi-dimensional integration in Eq.~(\ref{G0}) 
can be performed analytically following Ref.~\cite{Joyce:GreenElliptic:1998} leading to
 $\mathcal{G}_0(\varepsilon)=P(6/\varepsilon)/\varepsilon$, where 
\be\label{Pz}
P(z)=\frac{1-9\xi^4}{(1-\xi)^3(1+3 \xi)}\left [\frac{2}{\pi} Y(k_1)\right]^2.
\ee
Here $\xi$ and $k_1$ are functions of $z$ defined as  
\begin{eqnarray}\label{xi}
\xi(z)&=&\left (\frac{1-\sqrt{1-z^2/9}}{1+\sqrt{1-z^2}}\right)^{1/2},\\
k_1(z)^2&=&\frac{16 \xi^3}{(1-\xi)^3(1+3\xi)}, \label{k1} 
\end{eqnarray}
with $Y$ being the complete elliptic integral of the first kind. For a given disorder strength $W$,
the numerical band edges $\pm \varepsilon_{eb}(W)$ correspond to the energy values at which 
the imaginary part of the self-energy first vanishes, $\Im(\Sigma(\pm\varepsilon_{eb},W))=0$.

\bibliographystyle{apsrev}
\bibliography{biblio2bodynew.bib}

\end{document}